\documentstyle[aas2pp4]{article}

\newcommand{\stts}{4U~1626$-$67}

\begin{document}
\title{SIDEBANDS DUE TO QUASI-PERIODIC OSCILLATIONS IN \stts}

\author{Jefferson M.\ Kommers, Deepto Chakrabarty, 
  \& Walter H.\ G.\ Lewin}
\affil{Department of Physics and Center for Space Research,\\
  Massachusetts Institute of Technology, Cambridge, MA 02139\\
  {\bf To appear in ApJ Letters}\\ Contact:  {\tt kommers@space.mit.edu}}

\begin{abstract}
  The low-mass X-ray binary pulsar \stts\ shows 0.048 Hz
  quasi-periodic oscillations (QPOs) and red noise variability as well
  as coherent pulsations at the 0.130 Hz neutron star spin frequency.
  Power density spectra of observations made with the {\it Rossi X-ray
    Timing Explorer\/} show significant sidebands separated from the
  pulsar spin frequency (and its harmonics) by the QPO frequency.
  These show that the instantaneous amplitude of the coherent
  pulsations is modulated by the amplitude of the QPOs.  This
  phenomenon is expected in models such as the magnetospheric beat
  frequency model where the QPOs originate near the polar caps of the
  neutron star.  In the 4--8 keV energy range, however, the
  lower-frequency sidebands are significantly stronger than their
  higher-frequency complements; this is inconsistent with the
  magnetospheric beat frequency model.  We suggest that the 0.048 Hz
  QPOs are instead produced by a structure orbiting the neutron star
  at the QPO frequency.  This structure crosses the line of sight once
  per orbit and attenuates the pulsar beam, producing the symmetric
  (amplitude modulation) sidebands.  It also reprocesses the pulsar
  beam at the beat frequencies between the neutron star spin frequency
  and the QPOs, producing the excess variability observed in the
  lower-frequency sidebands.  Quite independently, we find no evidence
  that the red noise variability modulates the amplitude of the
  coherent pulsations.  This is also in contrast to the expectations
  of the magnetospheric beat frequency model and differs from the
  behavior in some high-mass X-ray binary pulsars.
\end{abstract}

% I selected keywords from the 1996 ApJ keyword list

\section{Introduction}
\label{sec:intro}
Quasi-periodic oscillations (QPOs) have been detected in at least 9
accretion-powered binary X-ray pulsars (see \cite{Take94} and
references therein; also \cite{Finger96}; \cite{Belloni90};
\cite{Zhang96}; \cite{Kommers97}).  Several physical mechanisms have
been proposed to explain the QPOs in X-ray binaries (see
\cite{Lewin88}; \cite{vdKlis95} for reviews).  One well-developed
model is the magnetospheric beat frequency model (MBFM) in which the
QPO centroid frequency ($\nu^{MBFM}_{QPO}$) represents the beat
frequency between the Keplerian orbital frequency at the inner edge of
the accretion disk ($\nu_K$) and the neutron star spin frequency
($\nu_s$): $\nu^{MBFM}_{QPO} = \nu_K - \nu_s$ (\cite{Alpar85};
\cite{Lamb85}).  Another model, which has received less discussion in
the literature, is the Keplerian frequency model (KFM; \cite{Bath74};
\cite{vdKlis87}).  In the KFM, inhomogeneities in the plasma orbiting
at the inner edge of the accretion disk modulate the X-ray intensity
by varying the optical depth along the line of sight; the QPO
frequency equals the Keplerian frequency at the inner edge of the disk
($\nu^{KFM}_{QPO} = \nu_K$).

\stts\ is a low-mass X-ray binary (LMXB) pulsar with a 7.67 s pulse
period (130 mHz spin frequency).  Despite extensive searches, no
Doppler shifts have been detected in the X-ray pulse arrival times
(\cite{Levine88}; \cite{Shinoda90}; \cite{Chakra98}).  The presence of
a low-mass companion star in a 42 minute prograde orbit around the
neutron star is inferred from photometric timing measurements on the
optical counterpart, KZ TrA, which shows pulsations at the same
frequency as the X-ray pulsations.  The power spectrum of the optical
variability shows additional weak pulsations in a sideband 0.4 mHz
below the main optical pulsation frequency (\cite{Middle81};
\cite{Chakra97}).  The interpretation is that the sideband represents
X-rays from the pulsar beam that have been reprocessed on the surface
of the companion star.  The 0.4 mHz shift to lower frequencies results
from the lower apparent pulsar frequency observed in a frame rotating
with the binary orbit of the companion (\cite{Middle81}).

The long-term behavior of the X-ray pulse frequency has been discussed
by Chakrabarty et al.\ (1997).  Although the pulsar was spinning up
($\dot{\nu_s} > 0$) for more than 13 years after its discovery, it
experienced a torque reversal in approximately June 1990.  It has been
spinning down ($\dot{\nu_s} < 0$) ever since (\cite{Chakra96}).  

QPOs at 48 mHz with fractional root-mean-square (rms) amplitudes as
high as 17\% have been seen in X-ray observations with {\it Ginga\/}
(\cite{Shinoda90}), {\it ASCA\/} (\cite{Angelini95}), and {\it RXTE\/}
(\cite{Chakra97}).  Chakrabarty (1998) recently detected these QPOs in
the optical $U$, $B$, and $R$ bands.

In this {\it Letter\/} we present evidence that neither the MBFM nor
the KFM provides a suitable explanation for the 48 mHz QPOs in \stts.
We propose instead that the QPOs are produced by some kind of coherent
structure (a ``blob'') orbiting the neutron star with an orbital
frequency equal to the QPO frequency, but the ``blob'' is not at the
inner edge of the accretion disk.

\section{Observations and Analysis}
\label{sec:obs}
\stts\ was observed with the {\it Rossi X-ray Timing Explorer\/} for a
total of 150 ks on 1996 February 10--15\footnote{These observations
  will be discussed further by Chakrabarty et al.\  (1998).}.  The
photon arrival times from the Proportional Counter Array (PCA) were
transformed to the solar system barycenter frame and binned at 0.125
s.  Two energy ranges were selected, 4--8 keV and 17--30 keV.  For
each energy range, the observations were partitioned into 400 s
intervals. Any intervals that contained data gaps or spacecraft slews
were omitted from further analysis.  The data for each of the 340
surviving intervals were flattened by subtracting the best-fit cubic
polynomial and then Fourier transformed to obtain a power spectrum.
An ensemble-averaged power spectrum for each energy range was obtained
by averaging the individual power spectra.  Uncertainties in the
average powers were computed from the sample variance of the
individual powers at each Fourier frequency.  These average power
spectra are shown in Figure \ref{fig:fullps}.

Several power spectral components can be identified in Figure 1.  The
feature at 48 mHz represents the QPOs identified previously (see
section \ref{sec:intro}).  The set of sharp peaks at integer
multiples of 130 mHz represents the harmonic structure of the coherent
pulsations (CPs).  A further red-noise (RN) component is apparent up
to $\sim 1$ Hz.  A local maximum in the RN component appears around 10
mHz, but this may be an artifact of the detrending procedure that was
applied before taking the Fourier transforms.

The finite length (400 s) of the Fourier transforms creates
troublesome side lobes around the CP peaks in the power spectrum (see
Figure \ref{fig:fullps}).  To suppress these and other complications
related to the presence of the coherent pulsations in the data, we
constructed a pulse-subtracted power spectrum.  For each 400 s data
segment, a fiducial pulse profile was obtained by folding the count rates
according to the frequency model given by Chakrabarty et al.\ (1997).
This pulse profile was subtracted from the data before taking the
Fourier transforms and constructing the average pulse-subtracted power
spectra.

For both energy ranges a best-fit model for the obvious RN, QPO, and
constant components of the pulse-subtracted power spectrum was found
by $\chi^2$ minimization.  A sum of two Lorentzian profiles was used
for the RN component and a single Lorentzian was used for the QPO
component.  The contributions of each component to the total best-fit
models are shown as dotted curves in Figure \ref{fig:fullps}.
Although this functional form provided a reasonable representation of
the gross features of the power spectra, the fits were formally
unacceptable.  The reduced chi-squared values were $\chi^2_\nu$ = 1.41
in the 4--8 keV range and $\chi^2_\nu$ = 1.29 in the 17--30 keV range
for 1526 degrees of freedom.  The data therefore support consideration
of more detailed models.

In the 4--8 keV range the best-fit model parameters for the QPO
component yield a fractional rms variability ($R$\/), corrected for
background, of $14.4 \pm 0.4$\%, a centroid frequency ($\nu_0$) of
$48.5 \pm 0.2$ mHz, and a FWHM ($\Gamma$) of $8.5 \pm 0.5$ mHz.  In
the 17-30 keV range the QPO parameters are $R = 22.4 \pm 0.5$\%,
$\nu_0 = 48.2 \pm 0.2$ mHz, and $\Gamma = 9.3 \pm 0.6$ mHz.

To look for more subtle features in the power spectra, we produced
residuals by subtracting the best-fit model from each pulse-subtracted
power spectrum.  Portions of these residuals below 600 mHz are shown
in Figure \ref{fig:residps}.  The arrows in Figure \ref{fig:residps}
show the positions where sidebands are expected if the QPO signal
modulates the amplitude of the coherent pulsations.  The origin of
these sidebands is the Fourier frequency-shifting theorem: if a
sinusoidal signal, $\cos(2 \pi \nu_s t)$, has a time-varying
amplitude, $f(t)$, then the Fourier transform of the source intensity
contains terms proportional to $F(\nu \pm \nu_s)$, where $F(\nu)$ is
the Fourier transform of $f(t)$ (for a complete discussion in the
context of the red-noise variability in X-ray pulsars see
\cite{Lazz97}; \cite{Burd97}).  If $f(t)$ represents the QPO signal
then its power spectrum $|F(\nu)|^2$ has a local maximum at the QPO
centroid frequency.  If $f(t)$ also modulates the instantaneous
amplitude of the coherent pulsations, the terms in the power spectrum
proportional to $|F(n \nu_s \pm \nu_{QPO})|^2$ produce symmetric
sidebands in the shape of the QPO peak around the frequencies $n
\nu_s$, where $n=1,2,\ldots$ is an integer.  Panel (a) of Figure
\ref{fig:residps} shows evidence for this effect, especially around
the second (390 mHz) and third (520 mHz) harmonics of the spin
frequency.  Panel (b) shows strong symmetric sidebands around the
fundamental (130 mHz).

A phenomenon that emits X-rays at the beat frequencies between the
pulsar harmonics and another oscillation (such as the QPOs) will
produce sidebands that appear either on the lower-frequency sides or
on the higher-frequency sides (but not both) of the harmonic
frequencies.  Panel (a) of Figure \ref{fig:residps} shows a clear
example of this.  The sidebands on the lower-frequency sides of the
fundamental (130 mHz) and the first harmonic (260 mHz) are much
stronger than the complementary sidebands on the higher-frequency
sides.  Additional oscillations in the source intensity must be
present at these beat frequencies (82 mHz and 212 mHz).

For greater sensitivity to the sidebands seen in Figure
\ref{fig:residps}, we produced harmonically folded residuals.  The
residuals from the neighborhood of each harmonic were shifted down to
the fundamental and summed.  Figure \ref{fig:foldres} shows the folded
residuals.  The dotted lines show the position of the pulsar
frequency.

We fitted independent Lorentzian profiles to each of the sidebands in
the folded residuals to measure the total rms variability contained in
them.  For the lower-frequency sideband in the 4--8 keV range, we
found $R = 5.7 \pm 0.4$\%, $\nu_0 = 83.2 \pm 0.6$ mHz, and $\Gamma =
8.8 \pm 0.2$ mHz; for the higher-frequency sideband, we found $R =
2.4 \pm 0.4$\%, $\nu_0 = 176.6 \pm 0.8$ mHz, and $\Gamma = 4.3 \pm
0.3$ mHz.  For the lower-frequency sideband in the 17--30 keV range,
we found $R = 8.3 \pm 0.5$\%, $\nu_0 = 83.2 \pm 0.7$ mHz, and $\Gamma
= 6.2 \pm 1.3$ mHz; for the higher-frequency sideband, we found $R =
7.5 \pm 0.5$\%, $\nu_0 = 177.6 \pm 0.6$ mHz, and $\Gamma = 5.5 \pm 1.2$
mHz.

We note that the folded residuals show no obvious sidebands in the
shape of the RN power spectrum components, consistent with an {\it
  absence\/} of amplitude modulation of the pulse intensity by the RN
variability.  In the 4--8 keV and 17--30 keV ranges, our fits to the
folded residuals place 90\%-confidence upper limits of 2.3\% and 3.1\%
(respectively) on the total rms variability contained in symmetric
sidebands due to amplitude modulation of the coherent pulses by the RN
variability.  These upper limits depend on the functional form used to
model the RN components, however.

\section{Discussion}
\label{sec:concl}
Significant sidebands are detected around the 130 mHz pulsar
frequency and its harmonics in power density spectra of 4U~1626$-$67.
The sidebands appear at frequencies $n \nu_s \pm \nu_{QPO}$, where $n
= 1,2, \ldots$ is an integer.  In the 17--30 keV range the sidebands
are mirror images of each other: they are symmetric in both frequency
and power amplitude.  We will refer to these as ``symmetric'' side
bands.  In the 4--8 keV range, however, the lower-frequency sidebands
contain significantly more power than the higher-frequency ones: they
are symmetric in frequency but {\it not} in power amplitude.  We will
assume that these ``asymmetric'' sidebands represent the
superposition of underlying symmetric sidebands {\it plus\/} some
additional power at the lower sideband frequencies ($n
\nu_s - \nu_{QPO}$).  We will refer to this excess power as ``enhanced
lower-frequency'' sidebands.

The presence of the symmetric sidebands suggests that the
instantaneous amplitude of the coherent pulsations contains a term
proportional to the 48 mHz QPO signal.  This phenomenon is expected in
models where some of the QPOs are produced near the polar caps of the
neutron star.  In the magnetospheric beat frequency model (MBFM), for
example, magnetically gated clumps of matter from the inner accretion
disk would follow the magnetic field lines onto the polar caps and
modulate the X-ray intensity there (\cite{Alpar85}; \cite{Lamb85}).
Symmetric sidebands could also occur in models such as the Keplerian
frequency model (KFM) where inhomogeneities in the accretion disk
quasi-periodically absorb some of the pulsar beam along the line of
sight.

The presence of the enhanced lower-frequency sidebands shows that the
situation is more complicated, however.  In the 4--8 keV range at
least two of the lower-frequency sidebands (the one at $\nu_s -
\nu_{QPO} = 82$ mHz and the one at $2 \nu_s - \nu_{QPO} = 212$ mHz)
are much stronger than their higher-frequency complements.  Since
amplitude modulation produces symmetric sidebands, there must be a
physical mechanism that produces additional variability at the
frequencies $n \nu_s - \nu_{QPO}$.  Neither the MBFM nor the KFM (as
usually formulated) provides a satisfactory explanation for both the
main QPO signal and the enhanced lower-frequency sidebands.

In the MBFM the QPOs represent a beat frequency between the Keplerian
frequency at the magnetopause and the pulsar frequency
($\nu_{QPO}^{MBFM} = \nu_{K} - \nu_{s}$).  If we set $\nu_K = 178$
mHz, the MBFM would predict the 48 mHz QPOs.  The presence of
symmetric sidebands would follow since the 48 mHz oscillations occur
in the accretion stream near the neutron star's polar caps.  But this
MBFM provides no mechanism that allows the QPOs to again ``beat''
against the rotating pulsar beam to produce the enhanced
lower-frequency sidebands at $\nu_s - \nu_{QPO} = 82$ mHz and $2 \nu_s
- \nu_{QPO} = 212$ mHz (\cite{Alpar85}; \cite{Lamb85}; \cite{Shib87}).

The KFM can explain the sideband structure (see our modified version
of this model below) but not the direct QPOs.  This is because the KFM
cannot explain QPOs with centroid frequencies {\it less\/} than the
neutron star spin frequency.  In the KFM the QPO frequency would be
the Keplerian frequency at the inner edge of the accretion disk (Bath
et al.\ 1974; \cite{vdKlis87}).  Since the QPO centroid frequency is
less than the pulsar spin frequency in \stts, the inner edge of the
disk would lie outside the co-rotation radius.  Centrifugal forces
exerted by the rotating neutron star magnetosphere would therefore
inhibit accretion (the ``propeller effect''; \cite{Ill75}).  The
radius corresponding to a 48 mHz Keplerian orbital frequency is $r_K =
1.3 \times 10^9$ cm, which exceeds the co-rotation radius $r_{co} =
6.5 \times 10^{8}$ cm.  The location of the inner edge of the disk is
less certain: depending on the distance to the source, $r_K$ could
exceed the Alfv\'en radius, $r_A \approx 3 \times 10^{9} d_{{\rm
    kpc}}^{-4/7}$ cm $\approx 9 \times 10^{8}$ cm, where we have
assumed a distance of 8 kpc and used the measurements by Orlandini et
al. (1997) of the 0.1--100 keV luminosity ($L_X = 6.6 \times 10^{34}$
erg s$^{-1}$ $d_{{\rm kpc}}^2$) and magnetic field strength ($B = 3.3
\times 10^{12}$ G).

If we dispense with the KFM's requirement that the QPO frequency
represents the {\it Keplerian frequency at the inner edge of the
  disk}, however, we obtain a scenario that can simultaneously account
for the direct QPOs, the symmetric sidebands, and the enhanced
lower-frequency sidebands.  Suppose a large coherent structure (a
``blob'') of material orbits the neutron star with a frequency roughly
equal to the QPO centroid frequency (48 mHz); this may or may not
represent the Keplerian frequency at the radius of the blob.  We use
the term ``blob'' to distinguish this structure from the ``clumps'' of
the MBFM (\cite{Lamb85}) and the ``inhomogeneities'' of the KFM (Bath
et al.\ 1974).  The direct QPO signal is produced as the ``blob''
modulates the optical depth to the accretion disk as it orbits.  Once
every orbit a portion of the ``blob'' crosses the line of sight
between the neutron star and the Earth and scatters X-rays from the
pulsar beam out of the line of sight.  This quasi-periodic attenuation
of the pulsar beam intensity produces the symmetric sidebands around
the spin frequency and its harmonics.

Suppose also that the ``blob'' orbits in the same sense as the pulsar
rotation, so it is illuminated by the pulsar beam with a frequency of
$\nu_s - \nu_{QPO} = 82$ mHz.  The first harmonic illuminates the
``blob'' with a frequency of $2\nu_s - \nu_{QPO} = 212$ mHz.  When the
``blob'' is not crossing the line of sight between the neutron star
and the Earth, it reprocesses X-rays from the pulsar beam and returns
some of them along the line of sight.  The reprocessed radiation
appears as oscillations at the frequencies of the enhanced
lower-frequency sidebands, $n \nu_s - \nu_{QPO}$.  The situation is
analogous to the reprocessing of the pulsar beam by the companion
star (\cite{Middle81}; \cite{Chakra97}).  The overall scenario is
similar to the production of orbital sidebands due to the modulation
of pulsations in intermediate polars (\cite{Warner86}).

\begin{figure}
\plotfiddle{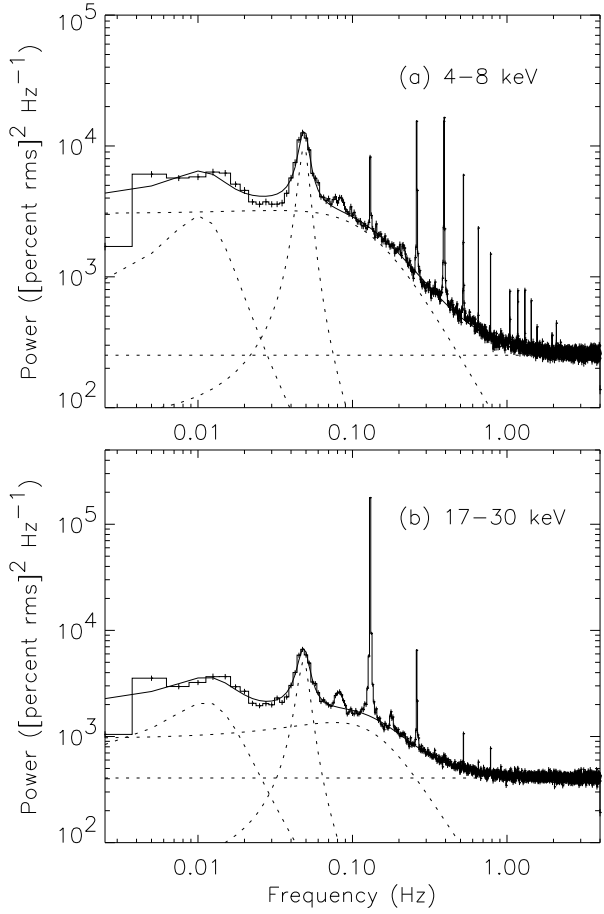}{5.0in}{0}{100}{100}{-144}{0}
%\plotone{rawps.eps}
\caption{Ensemble averaged power spectra of \stts\ (histogram).  The
  upper panel is from the 4--8 keV energy range, and the lower panel
  is from the 17--30 keV range.  The best-fit models for the
  pulse-subtracted power spectrum components are also shown (solid
  line).  The dotted lines show the contributions from each additive
  term in the best-fit models.
\label{fig:fullps}}
\end{figure}

\begin{figure}
\plotfiddle{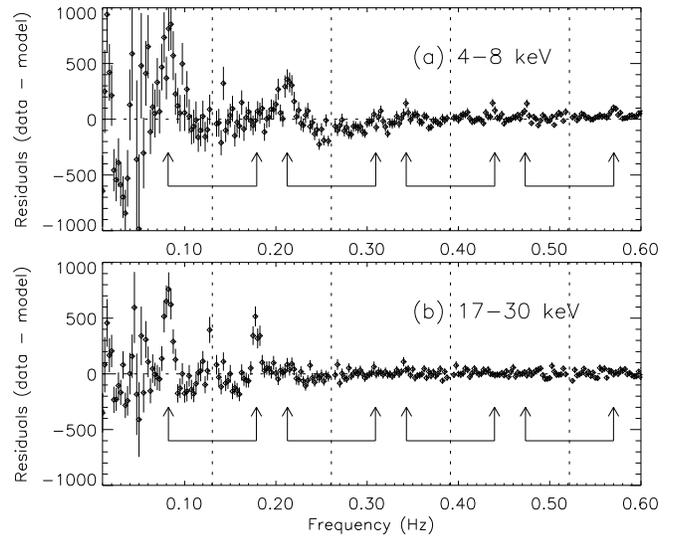}{6.0in}{0}{100}{100}{-144}{0}
%\plotone{resid.eps}
\caption{Residuals produced by subtracting the best-fit models from the
  pulse-subtracted power spectra.  Vertical dotted lines show the
  positions where the sharp peaks due to the coherent pulsations have
  been removed.  Pairs of arrows show the positions expected for
  symmetric sidebands if the QPO signal modulates the amplitude of
  the coherent pulsations.
\label{fig:residps}}
\end{figure}

\begin{figure}
\plotfiddle{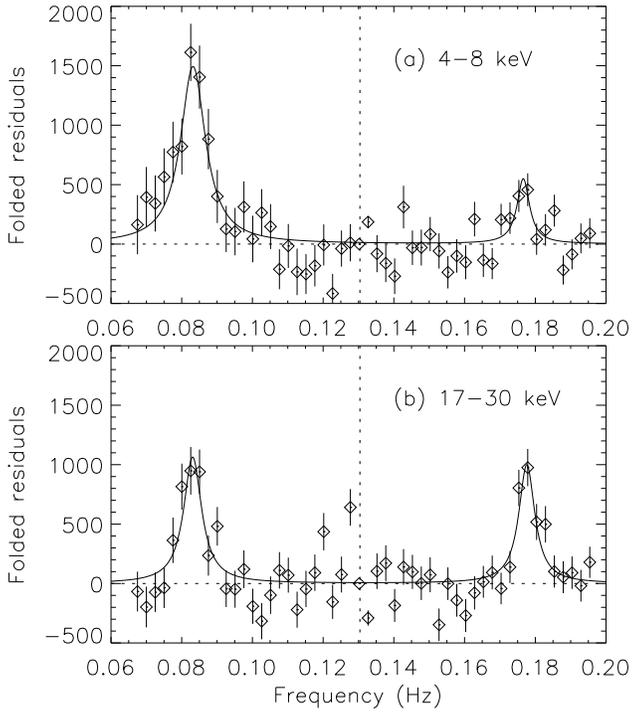}{5.0in}{0}{100}{100}{-144}{0}
%\plotone{foldres.eps}
\caption{Folded residuals (see text). Vertical dotted lines show the
  position where the harmonic peaks have been shifted to 130 mHz.  The
  solid line shows the best-fit Lorentzian functions to these
  residuals.
\label{fig:foldres}}
\end{figure}

The fact that the enhanced lower-frequency sidebands are found only
below the fundamental (130 mHz) and the first harmonic (260 mHz) in
the 4--8 keV range needs an explanation.  We suggest that this is
related to the reprocessing of the pulsar beam.  In the 17--30 keV
range the pulse profile is dominated by the fundamental and the first
harmonic (see Figure \ref{fig:fullps}).  If the ``blob'' reprocesses
hard X-rays and emits them at lower energies, the strengths of the
sidebands at 82 mHz and 212 mHz in the 4--8 keV range would reflect
the strengths of the corresponding pulse profile harmonics at {\it
  higher\/} energies (e.g.  17--30 keV).

The nature of the reprocessing ``blob'' is not clear.  It is unlikely
that there are many ``blobs'' scattered around the 48 mHz orbit,
because reprocessed emission from these would contribute with many
different phases and tend to wash out the oscillations responsible for
the enhanced lower-frequency sidebands.  On the other hand, it is not
clear why a single reprocessing structure of limited spatial extent
would survive in the accretion disk against the differential rotation
of nearby Keplerian orbits.  The QPOs appear to be stable on the time
scale of a decade or more, having been first observed in 1988
(\cite{Shinoda90}).

It is perhaps more likely that the orbital frequency of the ``blob''
is {\it not} Keplerian.  For example, the ``blob'' may represent a
superposition of oscillatory modes traveling as a stable wave packet
around the accretion disk.  Alpar \& Yilmaz (1997) have described
models of the normal-branch and horizontal-branch QPOs in LMXBs in
terms of wave packets of sound waves in the accretion disk.

To our knowledge the sideband structure has not been previously used
as a diagnostic of the QPO mechanism in X-ray pulsars.  Regardless of
whether the simple model we set forth above can stand up to closer
scrutiny, the sideband structure provides remarkably constraining
information on the possible QPO mechanisms.  An unusual sideband
structure was recently detected in the high-mass X-ray binary (HMXB)
pulsar Cen~X-3 (M.  Finger, private communication, 1997) that may
require a different interpretation than the one discussed here.

On the other hand, the sideband structure has been used to consider
the origin of RN variability in X-ray pulsars (\cite{Burd97};
\cite{Lazz97}).  If the RN is produced by inhomogeneities in the
accretion flow onto the magnetic poles of the neutron star, then the
instantaneous amplitude of the coherent pulsations should contain a
term proportional to the RN signal (just as for the QPO signal
discussed above).  For example, in the MBFM the same magnetically
gated clumps that produce the QPOs are expected to create a RN
component that represents the frequency content and lifetime
broadening of the shots produced by each clump (\cite{Lamb85};
\cite{Shib87}). The MBFM does not rule out the possibility of
additional sources of RN variability that are unrelated to the
magnetic gating mechanism, however.

We found no detectable modulation of the pulsar beam by the RN
component in \stts\ (see Figure \ref{fig:fullps}) even though the QPO
signal modulates the pulsar beam substantially.  The fractional rms
variability between 0.0025 Hz and $\sim 1$ Hz in the combined 4--8 keV
RN components was $42.5 \pm 0.5$\%; and that in the combined
17-30 keV RN components was $63.2 \pm 0.6$\%.  Yet the upper
limits on the total rms contained in the sidebands that would result
from a modulation of the coherent pulses by the RN are only $\sim
3$\%.  This suggests that most of RN does {\it not\/} originate near
the polar caps.  This is in contrast to case of the HMXB pulsars SMC
X-1, Vela X-1, 4U~1145$-$62, and possibly Cen X-3, in which the RN
variability does appear to modulate the coherent pulse amplitude
(\cite{Burd97}; \cite{Lazz97}).

To summarize, we have detected sidebands around the neutron star spin
frequency and its harmonics in \stts.  These show that the QPOs
modulate the amplitude of the coherent pulsations.  The presence of
the enhanced lower-frequency sidebands below the pulsar frequency and
its harmonics is inconsistent with the MBFM.  Interpreting the
lower-frequency sidebands as reprocessed radiation from the pulsar
beam, we have proposed a modification of previous (Keplerian
frequency) obscuration models that explains the observed QPOs and
sideband structure.  The strong RN component below 1 Hz does not
appear to significantly modulate the amplitude of the coherent
pulsations.

\acknowledgments We thank Mark Finger and Brian Vaughan for useful
discussions regarding the data analysis, and Dimitrios Psaltis and
Michiel van der Klis for comments on the manuscript.  J.\ M.\ K.\ 
acknowledges support from a NASA Graduate Student Researchers Program
Fellowship NGT8-52816.  D.\ C.\ was supported by a NASA {\it Compton
  GRO\/} Postdoctoral Fellowship under grant NAG5-3109.  W.\ H.\ G.\ 
L.\ acknowledges support from NASA.

% This gets ready for the bibliography
%\clearpage

% Here is the bibliography

\clearpage

%\clearpage

%\clearpage

\end{document}